\input harvmac
%\draftmode

\def \X {{\cal X}}
\def \II {{\cal I}}
\def \Tr {{\rm Tr}}
\def \tr {{\rm tr}}

\def\a{\alpha}

\def \const {{\rm const} }

\def \N {{\cal N}}

\def \k {\kappa}

\def \del {\partial}

\def \const {{\rm const}}
\def \ha{{\textstyle{1\over 2}}}

\def \a {\alpha}
\def \b {\beta}
\def \ov {\over}
 
\def\r {\rho}

\def \m {\mu}

\def \d {\delta}

\def \P {\Phi}

\def \inv {^{-1}}
\def \ov {\over }
\def \four{{\textstyle{1\over 4}}}
\def \fourth{{{1\over 4}}}

\def \d {\delta}

\def \lr { \lref}
\def \k {\kappa}

\def \four{{\textstyle{1\over 4}}}
\def \fourth{{{1\over 4}}}

\def \del {\partial}

\def \inv  {^{-1}}
\def \d {\delta} 
\def\a{\alpha} 
\def \b {\beta}
 
\def \ov {\over}

\def \half {{1 \ov 2}} 

\def\np {  {\it Nucl. Phys.} }
\def \pl { {\it  Phys. Lett.} }

\def \pr  { {\it Phys. Rev.} }

\lr\KT{I.R. Klebanov and A.A. Tseytlin, ``D-Branes and
Dual Gauge Theories in Type 0 Strings,'' {\tt hep-th/9811035}. }
\lr\KTU{I.R. Klebanov and A.A. Tseytlin,  unpublished (December, 1998).
}

\lr\GA{ M.R. Garousi, 
``String Scattering from D-branes in Type 0 Theories'',
{{\tt hep-th/9901085}}.}

\lr\ber{M. Bershadsky, Z. Kakushadze and
C. Vafa, 
``String expansion as large N expansion of gauge theories",
\np {\bf B523} (1998) 59, {\tt hep-th/9803076};
M. Bershadsky and A. Johansen,
``Large N limit of orbifold field theories,''
{\it Nucl. Phys.} {\bf B536} (1998) 141,
{\tt hep-th/9803249}.}

\lr\dm{M. Douglas and G. Moore, ``D-branes, quivers, and ALE
instantons,'' {{\tt hep-th/9603167}}.
}

\lr \DH { L. Dixon and J. Harvey, ``String theories in ten dimensions without space-time supersymmetry",  
{\it Nucl. Phys.} {\bf B274} (1986) 93;
 N. Seiberg and E. Witten,
``Spin structures in string theory", 
{\it Nucl. Phys.} {\bf B276} (1986) 272;
C. Thorn, unpublished.}

\lr\JM {J. Minahan, ``Glueball Mass Spectra and Other Issues for 
Supergravity Duals of QCD Models,'' {\tt hep-th/9811156}.  }

\lr \ferr{G. Ferretti and D. Martelli, ``On the construction of gauge theories from non critical type 0 strings,"
{\tt hep-th/9811208}. }

\lr\Malda {J. Maldacena, ``Wilson loops in large $N$
 field theories," {\it Phys. Rev. Lett.} {\bf 80} (1998) 4859, 
 {\tt hep-th/9803002};
S.-J. Rey and J. Yee, ``Macroscopic strings as heavy quarks in large $N$ gauge theory and anti-de Sitter supergravity", 
{\tt hep-th/9803001}.}

\lr\Sasha{A.M. Polyakov, ``String theory and quark confinement,''
{\it Nucl. Phys. B (Proc. Suppl.)} {\bf 68} (1998) 1, {{\tt hep-th/9711002}}. }

\lr\berg{O. Bergman and M. Gaberdiel, ``A Non-supersymmetric Open
String Theory and S-Duality,'' \np {\bf B499} (1997) 183,
{\tt hep-th/9701137}. }

%%%%%%%%%%%%%%%%%%%%%%%%%%%%%%%%%%%%%%%%%%%%%%%%%%%%%%%%%%%%
\lr  \kleb{
I.R. Klebanov, ``World volume approach to absorption by nondilatonic branes,''
  {\it Nucl. Phys.} {\bf B496} (1997) 231,
  {{\tt hep-th/9702076}}; 
S.S. Gubser, I.R. Klebanov, and A.A. Tseytlin, ``String theory and classical
 absorption by three-branes,'' {\it Nucl. Phys.} {\bf B499} (1997) 217,
  {{\tt hep-th/9703040}}.}

\lr  \gkThree{
S.S. Gubser and I.R. Klebanov, ``Absorption by branes and Schwinger terms in
  the world volume theory,'' {\it Phys. Lett.} {\bf B413} (1997) 41,
  {{\tt hep-th/9708005}}.}

\lr  \jthroat{
J.~Maldacena, ``The Large N limit of superconformal field theories and
  supergravity,'' {\it Adv. Theor. Math. Phys.} {\bf 2} (1998) 231, 
{{\tt
  hep-th/9711200}}.}

\lr  \US{
S.S. Gubser, I.R. Klebanov, and A.M. Polyakov, ``Gauge theory correlators
  from noncritical string theory,'' {\it Phys. Lett.} {\bf B428} (1998)
105,
  {{\tt hep-th/9802109}}.}

\lr  \EW{
E.~Witten, ``Anti-de Sitter space and holography,''
 {\it Adv. Theor. Math. Phys.} {\bf 2} (1998) 253, 
 {{\tt hep-th/9802150}}.}

\lr\GWP{D.J. Gross and F. Wilczek, {\it Phys. Rev. Lett.} {\bf 30}
(1973) 1343; H.D. Politzer, {\it Phys. Rev. Lett.} {\bf 30} (1973)
1346.}

\lr\GW{D.J. Gross and F. Wilczek, 
``Asymptotically free gauge theories 2," {\it Phys. Rev.}  {\bf D9} (1974) 
980. }

\lr  \AP{
A.M. Polyakov, ``The Wall of the Cave,''
{\tt hep-th/9809057.}}

\lr  \brane{
J.~Polchinski, ``Dirichlet Branes and Ramond-Ramond charges,'' {\it Phys. Rev.
  Lett.} {\bf 75} (1995) 4724,
{{\tt hep-th/9510017}}. }

\lr  \Jbook{
J. Polchinski, ``String Theory,'' vol. 2, Cambridge University Press,
1998.}

      \lr\TASI{
J. Polchinski, ``TASI Lectures on D-Branes,''
{\tt hep-th/9611050.}  }

\lr  \Witten{
E.~Witten, ``Bound states of strings and p-branes,'' {\it Nucl. Phys.} {\bf
  B460} (1996) 335, {{\tt
  hep-th/9510135}}.  }

\lr  \hsdgt{
G.T. Horowitz and A.~Strominger, ``Black strings and p-branes,'' {\it Nucl.
  Phys.} {\bf B360} (1991) 197;
M.J. Duff  and J.X. Lu, 
``The selfdual  type IIB  superthreebrane,"
{\it Phys. Lett.}  {\bf B273} (1991)  409.}
\lr\GT{
 G.W. Gibbons and  P.K. Townsend,
``Vacuum interpolation in supergravity via super p-branes",
 {\it Phys. Rev. Lett.} {\bf 71} (1993) 3754, 
{\tt hep-th/9307049}.
}

\lr\KW{I.R. Klebanov and E. Witten, ``Superconformal field theory on
threebranes at a Calabi-Yau singularity,'' {\tt hep-th/9807080};
S.S. Gubser, ``Einstein manifolds and conformal field theories,''
{\tt hep-th/9807164}.}

\lr\GNS{
S.S. Gubser, N. Nekrasov and S. Shatashvili,
``Generalized conifolds and four dimensional ${\cal N}=1$
superconformal theories,''
{\tt hep-th/9811230}.
}

\lr\Gir{L. Girardello, M. Petrini, M. Porrati, and  A. Zaffaroni, 
``Novel Local CFT and Exact Results on Perturbations of N=4
   Super Yang Mills from AdS Dynamics", 
{\tt hep-th/9810126};
J. Distler and F. Zamora, 
``Non-Supersymmetric Conformal Field Theories from Stable
   Anti-de Sitter Spaces", 
{\tt hep-th/9810206}.}

\lref \tat {A.A. Tseytlin, 
``On the tachyonic terms in the string effective action", 
{\it Phys Lett.}  {\bf B264} (1991) 311.}
\lref \bank{T. Banks, ``The tachyon potential in string theory",
{\it Nucl. Phys.} {\bf B361} (1991) 166.}

%C. Angelantonj, ``Nontachyonic Open Descendants of the 0B String Theory",
%{\tt  hep-th/9810214}.  }
\lr\mig{A.A. Migdal, ``Hidden Symmetries of Large N QCD,"
{\it Prog. Theor. Phys. Suppl.} {\bf 131} (1998) 269, {\tt hep-th/9610126}.}
  \lr\alva{E. Alvarez, C. Gomez and T. Ortin,
``String representation of Wilson loops", 
{\tt hep-th/9806075}. }

 \lr\jones{D.R.T. Jones,
``Asymptotic behavior  of supersymmetric Yang-Mills theories
in the two-loop approximation," {\it Nucl. Phys.} {\bf  B87}
 (1975) 127; ``Charge renormalization 
in a supersymmetric Yang-Mills theory,"
{\it  Phys. Lett.}  {\bf B72} (1977) 199; 
E. Poggio and H. Pendleton,  
``Vanishing of charge renormalization 
and anomalies in a supersymmetric gauge theory,''
 {\it  Phys. Lett.}  {\bf B72}  (1977) 200.}

\lr\MV{ 
M.E. Machacek and M.T. Vaughn, 
``Two-loop renormalization group equations in a general 
quantum field theory I: Wave function renormalization,"
{\it Nucl. Phys.} {\bf B222} (1983) 83;
``II. Yukawa couplings'', \np  {\bf B236} (1983) 221;
``III. Scalar quartic  couplings'', \np  {\bf B249} (1985) 70.
}

\lr\PW{A. Parkes and P. West, 
``Finiteness in rigid supersymmetric theories", 
\pl B138 (1984) 99.}

\lr\baz{T. Banks and A. Zaks,  ``On the
phase structure of vector-like gauge theories with massless
fermions,"
 {\it Nucl. Phys.} {\bf B196} (1982)  189. } 
\lr\KTnew{
I.R. Klebanov and A.A. Tseytlin, ``Asymptotic Freedom and
Infrared Behavior in the Type 0 String Approach to Gauge Theory,'' 
{\tt hep-th/9812089}. }

\lr \KS  {S.~Kachru and E.~Silverstein, ``4d conformal field theories
and strings on orbifolds,''  {\it Phys. Rev. Lett. } {\bf 80} (1998)
4855, 
{{\tt hep-th/9802183}}.}
\lr\LNV {A.~Lawrence, N.~Nekrasov and C.~Vafa, ``On conformal field
theories in four dimensions,'' {\it Nucl. Phys.} {\bf B533} (1998) 199, 
{{\tt hep-th/9803015}}.}

\lr \FR { P. Frampton, ``ADS/CFT String Duality and Conformal Gauge
Theories,'' {\tt hep-th/9812117}.}
\lr \DB  {J. Blum and   K.  Dienes, ``Strong / Weak Coupling Duality
Relations for
Non Supersymmetric String Theories",  {\it Nucl. Phys.} {\bf  B516}  
(1998) 83, 
{\tt hep-th/9707160.} } 

\lr\PW{A. Parkes and P. West, 
``Finiteness in rigid supersymmetric theories", 
{\it  Phys. Lett. } {\bf B138}  (1984) 99.}

\lr\JS {J.H. Schwarz, ``Covariant field equations
 of chiral N=2,D=10 supergravity", \np {\bf B226} (1983) 269.  } 

\lr\bisa{
M. Bianchi and  A. Sagnotti,
``On the Systematics of Open String Theories",
{\it Phys. Lett.} {\bf B247}  (1990) 517;
A. Sagnotti, ``Some Properties of Open - String Theories", 
{\tt hep-th/9509080}; ``Surprises in Open-String Perturbation Theory", 
 {\it Nucl.Phys.Proc.Suppl.} {\bf  B56}  (1997) 332, 
{\tt hep-th/9702093}.}

\lr \CT{M. Cveti\v c   and A. A. Tseytlin, 
``Charged string solutions with dilaton and modulus fields'',
{\it Nucl. Phys.} {\bf B416} (1994) 137, {\tt hep-th/9307123}.}

\lr \MT{T. Banks  and  M.B. Green,  
 ``Nonperturbative effects in   $AdS_5 \times   S^5$ string theory  
 and d = 4 susy  
 Yang-Mills",  {\it J. High Energy Phys.}  9805 (1998)  002, 
 {\tt  hep-th/9804170 };
R.R. Metsaev, A.A. Tseytlin, ``Type IIB superstring action in $ AdS_5
\times  S^5$  background'', {\it Nucl. Phys.} {\bf B533}
(1998) 109, {\tt  hep-th/9805028};
R. Kallosh and A. Rajaraman,
``Vacua of M theory and string theory'', 
{\it Phys. Rev.} {\bf D58} (1998) 125003, {\tt hep-th/9805041}.} 

\lr\BF{P. Breitenlohner and D.Z. Freedman,
``Stability in gauged extended supergravity'',
{\it Ann. Phys.} {\bf 144} (1982) 249.}

\lr\DS{
M. Dine, P. Huet and N. Seiberg, 
``Large and small radius in string theory'',
{\it Nucl. Phys.}   {\bf B322} (1989) 301;
 R. Rohm,
``Spontaneous supersymmetry breaking in supersymmetric field theories'',
 \np {\bf B237}  (1984) 553.}

\lr\br{M. Berkooz and S.-J. Rey, ``Nonsupersymmetric stable vacua of
M theory,'' {\tt hep-th/9807200}.}

\lr\AW{J.J. Atick  and E. Witten, 
``The Hagedorn transition and the number of degrees of freedom 
of string theory'', 
\np {\bf B310} (1988) 291.}

\lr\newWit{
E. Witten, ``Anti-de Sitter space, thermal
phase transition, and confinement in gauge theories,''
{\it Adv. Theor. Math. Phys.} {\bf 2} (1998) 505, {\tt hep-th/9803131.}
}

\lr\JO{I. Jack and H. Osborn, ``General background field calculations with fermion fields", 
\np {\bf B249} (1985) 472.}
\lr \OO{T.P. Cheng, E. Eichten and L.-F. Li, 
``Higgs phenomena in asymptotically free gauge theories,"
\pr {\bf D9} (1974) 2259.}

\lr\NS{N. Nekrasov and S. Shatashvili,
``On non-supersymmetric CFT in four dimension,''
{\tt hep-th/9902110.}
}

%%%%%%%%%%%%%%%%%%%%%%%%%%%%%%%%%%%%%%%%%%%%%%%%%%%%%%

\baselineskip8pt
\Title{\vbox
{\baselineskip 6pt
{\hbox {PUPT-1833}}{\hbox{Imperial/TP/98-99/37 }}
{\hbox{hep-th/9901101}} 
{\hbox{   }}
}}
{\vbox{\vskip -30 true pt
\centerline {A Non-supersymmetric Large N CFT}
\medskip
\centerline {from Type 0 String Theory}
\medskip
\vskip4pt }}
\vskip -20 true pt 
\centerline{ Igor R. Klebanov}
\smallskip\smallskip
\centerline{Joseph Henry Laboratories, Princeton University, 
Princeton, New Jersey 08544, USA}
\bigskip
\centerline  {Arkady A. Tseytlin\footnote{$^{\dagger}$}{\baselineskip8pt
Also at  Lebedev  Physics
Institute, Moscow.} }
\smallskip\smallskip
\centerline {  Blackett Laboratory, Imperial College,  London SW7 2BZ, U.K.} 

\bigskip\bigskip
\centerline {\bf Abstract}
\baselineskip10pt
\noindent
\medskip
We show that type 0B theory has a classical $AdS_5\times S^5$
solution and argue that it is stable at the string-theory level
for small enough radius. The dual 4-d conformal field theory 
is the infrared limit
of the theory on $N$ electric D3-branes coincident with $N$
magnetic D3-branes. We explicitly construct this 
$SU(N)\times SU(N)$ gauge theory with global
$SO(6)$ symmetry and 
verify that the one-loop term in the beta function vanishes exactly,
while the two-loop term  vanishes in the large $N$ limit.
We find that this theory is related by a certain projection to the 
maximally supersymmetric Yang-Mills theory
suggesting its large $N$ conformal invariance to all orders
in perturbation theory.

\bigskip
 
%%%%%%%%%%%%%%%%%%%%%%%%%%%%%%%%%%%%%%%%%%%%%%%%%%%%%%%%%
\Date {January 1999}

%%%%%%%%%%%%%%%%%%%%%%%%%%%%%%%%%%%%%%%%%%%%%%%%%%%%%%%%%%%%%%%%%%%
\noblackbox \baselineskip 15pt plus 2pt minus 2pt 
%\baselineskip 20pt plus 2pt minus 2pt

%%%%%%%%%%%%%%%%%%%%%%%%%%%%%
\newsec{Introduction}
%%%%%%%%%%%%%%%%%%%%%%%%%%%%%%%%%

In recent papers \refs{\KT,\JM,\KTnew} non-supersymmetric
large $N$ gauge theories dual to certain backgrounds of
the type 0 string theory \DH\ were studied. 
This work was inspired by the recently discovered relations between
type II strings and superconformal gauge theories on $N$ coincident
D3-branes \refs{\kleb,\gkThree,\jthroat,\US,\EW}, and, in particular, 
by Polyakov's suggestion \refs{\AP} (building on his earlier work \Sasha) 
that the type 0 string theory in dimensions $D \leq 10$ is a natural
setting for extending this duality to {\it non}-supersymmetric 
gauge theories.

The type 0 string \DH\ has world sheet supersymmetry
%,\foot{
%Possible relevance of world sheet supersymmetry to string description of
%gauge theories was also advocated in \refs{\Sasha,\mig,\alva}.}
but no space-time supersymmetry. In both type 0A and type 0B theory
the fermions are completely projected out of the spectrum.
The massless bosonic fields are  as in the corresponding 
type II theory (A or B), 
but with the  doubled set of Ramond-Ramond (R-R)  fields. 
The spectrum of
type 0 theory also contains a tachyon,  but, as suggested in 
 \refs{\AP,\KT}, its presence 
does not spoil the application to large $N$ gauge theories.
In fact, the tachyon background provides a natural mechanism for
breaking the conformal invariance and introducing the
renormalization group flow \refs{\KT,\JM,\KTnew}.

In \KT\ the $3+1$ dimensional $SU(N)$ theory coupled to 6 adjoint massless
scalars was constructed as the low-energy description of $N$
coincident {\it electric} D3-branes. The conjectured dual type 0B
background thus carries $N$ units of electric 5-form flux.
The set of equations for a background with a 5-form flux
follows from the following action:
 $$ 
S_0= - { 1 \ov 2 \k^2_{10}} \int d^{10} x \sqrt{ -g} 
\bigg[ R - \half \del_n \P \del^n \P
$$ \eqn\fii{ 
-\  \fourth (  \del_n  T  \del^n T   + m^2 e^{\half \P}  T^2) 
 - \   { 1 \ov 4 \cdot 5!}  f(T)  F_{n_1...n_5} F^{n_1...n_5} + 
\ldots \bigg] \ , }
where  
$g_{mn}$ is the Einstein-frame metric, \ $  m^2 = - {\textstyle{ 2 \ov \a'} }$,
and \KT\ 
\eqn\cf{ f(T)= 1+ T +   O(T^2)\ .}
In \refs{\KT,\JM,\KTnew} it was shown that, if
the tachyon--R-R field coupling function $f(T)$ has a minimum,
then there exists an asymptotically free type 0 background
with the tachyon  approximately localized near the minimum. 
The tachyon background induces a radial variation of the 
dilaton $\Phi$ which has the physical interpretation of renormalization
group flow.
The Einstein metric  is
approximately $AdS_5\times S^5$, while the effective gauge coupling
evolves logarithmically with the scale \refs{\JM,\KTnew}.
In \refs{\JM,\KTnew} the infrared behavior of the background was
also considered. It turns out that there
is a fixed point at infinite coupling which,
just like the fixed point at zero coupling, is characterized by
the $AdS_5\times S^5$ Einstein frame metric \KTnew.
We argued that there is a  renormalization group 
trajectory extending all the way from the
zero coupling fixed point in the UV to the infinite coupling fixed
point in the IR.

In this paper we instead consider a background of type 0 theory which
has exact conformal invariance: in fact, a line of fixed points.
The existence of such a background was pointed out already
in \KT: it is the near-horizon region of the self-dual
3-brane of type 0B theory which, just as in the type IIB case 
\refs{\GT}, turns
out to be $AdS_5\times S^5$. In the type IIB case the dual
gauge theory is ${\cal N}=4$ SYM  \refs{\jthroat} but in the type 0B case it is
different, and its investigation is the main purpose of this
paper. 

The self-dual 3-brane is constructed by stacking equal numbers
$N$ of coincident electric and magnetic 3-branes.
We will show that the non-supersymmetric 
theory on this stack is the $SU(N)\times SU(N)$
gauge theory coupled to 6 adjoint scalars of the first $SU(N)$, 
6 adjoint scalars of the second $SU(N)$, and fermions in the 
{\it bifundamental} representations -- 4 Weyl fermions in the
$({\bf N}, \overline {\bf N})$ and 4 Weyl fermions in the
$(\overline {\bf N}, {\bf N})$. We will argue that
this theory is conformal in the large $N$ limit. Since it has the
maximal global $SO(6)$ invariance, this theory is not of the kind that
occurs on the D3-branes of type IIB theory
placed at the orbifold singularity of $R^6/\Gamma$
%it does not belong 
%to a class of $\N=0$ large $N$  `orbifold' CFT's where
%the orbifold group is a subgroup of the $SO(6)$ R-symmetry of the
%${\cal N}=4$ theory 
\refs{\dm,\KS,\LNV,\ber} (the latter theories are dual to
the $AdS_5\times S^5/\Gamma$ backgrounds of type IIB theory
\refs{\KS,\LNV}). 

In section 2 we consider the classical equations for a dyonic type 0B 
3-brane and find that for equal electric and magnetic charges
there is an $AdS_5\times S^5$ near-horizon solution.
In section 3 we discuss the dual field theory and show that the
1-loop beta function vanishes exactly, while the 2-loop correction
vanishes in the large $N$ limit. We further argue that all higher
order terms vanish in the large $N$ limit because the field theory 
can be interpreted as
a $Z_2$ projection of the ${\cal N}=4$\  $U(2N)$\  SYM theory 
by $(-1)^F$ combined with a $Z_2$ gauge group twist.
In section 4  we discuss
the analogy  between  this projection  and a 
similar $(-1)^F$ projection 
which connects the type II and type 0 string theories 
compactified on a circle
\refs{\DH,\AW,\DS,\DB}.

%%%%%%%%%%%%%%%%%%%%%%%%%%%%%%%%%%%%%%%%%%%%%%%%%%%%%%
\newsec{The $AdS_5\times S^5$ background of type 0B theory}
%%%%%%%%%%%%%%%%%%%%%%%%%%%%%%%%%%%%%%%%%%%%%%

Lets us parametrize the 10-d string-frame and Einstein-frame  metric 
as in \KT \ ($\m=0,1,2,3$ are the 4-d indices) 
\eqn\met{
ds^2 = e^{\half \P} ds^2_E\ , \ \ \ \ \ \ \ 
ds^2_E = e^{\half \xi - 5 \eta} d\r^2  + e^{-\half \xi}  dx^\m dx^\m
 + e^{\half \xi -  \eta} d\Omega^2_5 \ . }
%where $\tilde \Phi= \Phi- \Phi_0$ and $\Phi_0$ is the background value
%of the dilaton field which determines the string coupling constant,
%$g_{\rm s}= e^\Phi_0$.
In terms of these variables
the effective action describing radial evolution of the fields
corresponding to \fii\ becomes \KT
\eqn\eas{  
S=  \int d\r \bigg[ \ha  \P'^2  + \ha  \xi'^2   
- 5   \eta'^2 + \four  T'^2
   + U (\P,\xi,\eta,T) \bigg] \ , 
}
\eqn\eio{
U = -  \ha  T^2  
e^{{1 \ov 2} \P + {1 \ov 2}  \xi   - 5 \eta }
   -  20   e^{-4\eta }   \  +  \  h (T)\  e^{-2\xi}  \ . } 
Here $\a'=1$,  
and  $\P, \xi, \eta$  and $T$ are functions of $\r$. For the case  of the electric 5-form background, it was shown in
\KT\ that $h(T)= Q^2   f^{-1} (T)$, where  $f(T)$ is the tachyon--5-form
coupling function \cf\ and $Q$ is the electric charge. 
Generalizing to the presence
of magnetic 5-form charge $P$ is straightforward,  and we find that
\eqn\duw{ h(T) = P^2 f(T) + Q^2   f^{-1} (T)\ .
}
As in the familiar dilaton-vector coupling case, the potential that 
originates from the $f(T) F_5F_5$ term in the 
full $D=10$ action is invariant under
the electro-magnetic duality\ \  $P\leftrightarrow Q, \ f \leftrightarrow
f\inv$.

The function $ h(T)$ has an extremum at $T=T_{0}$ such that 
$f(T_0) = Q/P$ (we assume that $f(T), Q, P$ are non-negative). 
For $Q=P$ ($\sim N$)  we  thus  get  $f(T_0)=1$ which has a solution
$T_0=0$ (cf. \cf).
Then the tachyon equation is solved by $T=0$, 
the dilaton equation is  solved by $\Phi=\Phi_0$,
and the equations for $\xi,\eta$ 
are solved by the usual $AdS_5 \times S^5$  background
(note that  $h(0) = 2 Q^2$ for $P=Q$) 
\eqn\sios{
\xi  =   \ln (2Q) + \ln  \r  \ , \ \ \ \  \ 
\eta =  \ln 2 +  \ha  \ln  \r \ .  }
This  is the  near-horizon ($2 Q e^{\Phi_0}  \r \gg 1$) 
 limit of the full  3-brane metric. After the change of variables
$\r= u^{-4}$, the $AdS_5 \times S^5$ metric is 
\eqn\ads{ ds^2_E=  R^2_0  ( {du^2\ov u^2}  + {u^2\ov 2 R^4_0} dx^\m dx^\m 
+ d \Omega^2_5  ) \ , \ \ \ \ \ \ \ \   
 R^2_0 = \ha  (2Q)^{1/2}\ .  } 
The  4-th power of the radius $R$ of  the corresponding string-frame metric, 
i.e. $\ha e^{\Phi_0} Q$,  is proportional to the `t Hooft
coupling $N g_{\rm YM}^2$ 
of the gauge theory on the stack of $N$ electric D3-branes
and $N$ magnetic D3-branes.

An important issue is that of stability of the 
background  (cf. \refs{\BF,\br}).
Considering small perturbations of the $AdS_5 \times S^5$
background in the $T$-direction\foot{The perturbations of the
scalar $T$ near $T=0$ do not mix with 
the dilaton-graviton  perturbations since the background value
of $F^2_5$ is zero in the self-dual 3-brane case.
In general, the mixing of $T$ and the R-R potential $C_4$
fluctuations is already effectively accounted for in $U$
(see \CT\ for a  discussion of stability of a similar  
dilaton-vector system in 4 dimensions).}
we learn that they are no longer tachyonic for sufficiently small 
$Q e^{\Phi_0}$.
Indeed, the leading  quadratic terms in the fluctuation
of $T$ near $T_0=0$ in the potential $U$ become:
\eqn\eiii{
U_2 = { 1 \ov 4 \rho^2}  \bigg[ - { 1 \ov 16}     (2 Q e^{\P_0}  )^{1/2} 
  +   f'^2 (0) \bigg]  T^2  + O(T^3) \ . }
Thus, the background is stable if
$16 f'^2 (0) \   \geq   \       (2 Qe^{\P_0})^{1/2},$  
which is satisfied for small enough $Qe^{\P_0} \sim N g_{\rm s}$
since $f'(0)=1$. More precisely,
the  stability condition for a scalar field with mass
$M^2$ in $AdS_5$ space-time  with radius $R$ is  \BF:
 $ M^2 \geq -  4R^{-2}$, i.e.
in the present case $M^2 \geq  - 8 (2 Qe^{ \P_0})^{- 1/2}$.
Here the effective mass parameter (measured in the string frame, $\a'=1$) is 
$M^2 = -2  + 32   f'^2 (0) (2 Qe^{ \P_0})^{-1/2}$,
so the stability  condition  is actually 
\eqn\stab{ 4  + 16 f'^2 (0) \geq  (2 Qe^{\P_0})^{1/2} \ .
}
Note that,  once one is allowed to consider the
$AdS_5$ radius to be of order of $\sqrt{\a'}$, one in principle does not
need the $f(T)$ coupling to achieve the stability of the string 
tachyon in the
$AdS_5 \times S^5$ background. However,
the coupling function $f(T)$ {\it is}
non-trivial in type 0B theory, and since $f'(0) = 1$ \KT, 
its contribution to  the above  condition
is considerable. Explicitly, we get
$ \ha Qe^{\P_0} = 4 \pi  g_{\rm s}  N  \a'^2 =  g^2_{\rm YM} N $,
and thus the following stability condition on the 't Hooft coupling:\ \ 
$g^2_{\rm YM} N  < 100$.    

The stability
is related to the fact that the $T=0$ solution is an attractor:
a generic solution with $T=T^{(0)}$ at $\r=0$ evolves to 
the $T=0$ solution at large $\r$. This is easy to see
explicitly, e.g., by approximating $f(T)$ by $e^T$ 
which is equal to $f_2(T)= 1+ T + \ha T^2$  
at small $T$.
Then $h(T) = 2 Q^2 \cosh T$ which indeed gives
a positive shift of  the coefficient of the $T^2$ term in the potential.
Let us note in passing  that the choice $f(T)=e^T$, which is in
agreement with
the perturbation-theory  result $f_2(T)$ of \KT\ at small $T$
may have more general significance.
It is a natural definition of the R-R -- tachyon coupling 
consistent with the expected  electric--magnetic
5-form duality of the type 0B theory. 
In fact, consideration of the D-brane action couplings in \KT\ implies 
that the sign of the tachyon tadpole should be changed  when changing
from the electric to the magnetic 3-brane. Thus, if we require that
under the 5-form duality $T\rightarrow - T$, then the R-R -- tachyon
coupling is fixed to be $e^T F_5 F_5$. It is not completely clear,
however, whether the massless field/tachyon effective action
has to be symmetric under the duality.
%That implies that the electric and magnetic parts of the R-R
%strengths should couple  to $T$ as $ e^T F^2_{el} + e^{-T} F^2_{mag}$.

A stability criterion similar to \stab\ was already found in \KT\ 
where it was noted that it is opposite to the condition of validity of the
gravity (leading order in $\a'$) approximation.
Indeed, based on \stab\ one may hope that the theory is stable for
$Qe^{\P_0} < O(1)$, but here the curvature of the background is of
order 1 in string units and it is not clear { a priori} 
whether the effective gravity approximation can be trusted in the first place.
We would like to argue, though, that it indeed can be trusted
due to the special nature of the
$AdS_5\times S^5$ solution.

As we explained in \KT, the tree-level 
effective action of the type 0B theory coincides 
with that of the type IIB theory on the subsector of bosonic  fields
%that appears in
of  the type IIB theory.
Since the present self-dual 
solution has $T=0$ and $F_5= \tilde F_5$ (i.e. the non-vanishing 
fields belong to the spectrum of type IIB theory), we are 
in a position to argue that our solution is exact to all orders in $\a'$
just as it is in type IIB theory \MT.
The stability condition is also 
expected to hold (at least qualitatively)
since the higher-order in $\a'$ corrections
to the $T^2$ term in the effective action 
are likely to involve powers of 
the Weyl tensor, $F^2_5$ and the Ricci scalar which all vanish
in the $AdS_5\times S^5$ background.
%\foot{While for a generic
%2-d CFT, e.g., a WZW  model, 
% one certainly  does not expect the anomalous dimension 
%of the tachyon operator not to receive $\a'$ (or $1/k$)  corrections,
%this may happen in the special cases of `balanced' 
%combinations, like equal-level 
%$SL(2) \times SU(2)$ or $AdS_3 \times S^3$, 
%which also has  zero  total Ricci scalar 
%and Weyl tensor.}
If one accepts this argument, then one is led to the conclusion that
the dual field theory which lives on the stack of D3-branes
is a stable large $N$ CFT as long as the `t Hooft coupling
$N g_{\rm YM}^2$ is not very large: 
less than a number of order $10^2$.
In the next section
we present a weak-coupling analysis that provides some
support for this conclusion.

We conclude that  
while the flat space vacuum is unstable 
in type 0  string theory, the $AdS_5\times S^5$ with self-dual
5-form flux should be a stable background for sufficiently small radius. 
The R-R charge works to stabilize the tachyon as proposed in \KT.
Our conclusions are independent of the details
of the tachyon -- R-R coupling 
function $f(T)$: any function such that $f(0)=1$ and $f'(0)=1$
leads to the stable $AdS_5\times S^5$ background. 
These properties of $f(T)$ are obviously satisfied by \cf, and 
therefore our conclusions appear to be robust.

%%%%%%%%%%%%%%%%%%%%%%%%%%%%%%%%%%%%%
\newsec{Analysis of the Field Theory}
%%%%%%%%%%%%%%%%%%%%%%%%%%%%%%%%

The GSO projections for open strings connecting the D-branes of
type 0 theory were considered in \refs{\berg,\KT}.\foot{Open  string
 descendants of type 0B theory
 were  constructed  by orientifold projection in \bisa.} 
In considering parallel like-charged D-branes we find the same
bosonic spectrum as in the corresponding type II theory,
but the fermions are excluded. Thus, for 
$N$ parallel electric D3-branes we
get the $U(N)$ gauge theory coupled to 6 adjoint scalars \KT.
Exactly the same theory is found in the case of  
$N$ parallel magnetic D3-branes. Now we would like to
stack $N$ parallel electric and $N$ parallel magnetic D3-branes.
The new and somewhat surprising feature is that an open string
connecting an electric and a magnetic D3-brane is actually a {\it fermion}
\berg. It is clear that these fermions will transform in the
bifundamental representations of the $U(N)\times U(N)$.
It is remarkable that, although type 0 theory has no fermions 
in the bulk, they are present on  the dyonic  D3-brane.

It follows  already from the perturbative analysis that a composite
`self-dual' type 0B 
D3-brane consisting of an electric D3-brane coincident with a
magnetic D3-brane has a number of special properties. From the results of
\KT\ one finds that the tachyon tadpole has an opposite
sign for the magnetic D3-brane compared to the electric one;
therefore the tadpole cancels on
the self-dual D3-brane.\foot{Indeed, the resulting self-dual
D3-brane action is then decoupled from
the tachyon and from 
 the second  R-R potential $\bar C_4$  with
{\it anti}-selfdual $F_5$    
\KT, and so is essentially the same as in type IIB.
In general,  the BI part of the  action is
$\int d^4 x (1 + \four  q \bar q T+...)  e^{-\Phi} \sqrt{-\det( G_{\a\b} +
\del_\a X \cdot \del_\b X + 2\pi F_{\a\b} )},$
where $q=\bar q=1$ for the electric, $q=-\bar q=1$ for the magnetic
and $q=2, \bar q=0$ for the self-dual D3-brane.
The  coefficient of the tachyon tadpole was found in \KT, 
and  then on reparametrization invariance (and $T$-duality) 
grounds one expects 
that the tachyon function $k(T)= 1 + \four  q \bar q T + O(T^2)$  
multiplies the $\sqrt {-\det (G + F)}$ in the BI action. 
We have checked this by  directly computing the coefficient of the 
$T F_{\a\b} F^{\a\b}$ term from the corresponding disk diagram.
This calculation was independently carried out in 
the recent paper \GA\  where  
it was also conjectured and then verified that 
the tachyon coupling function must appear in front
of the square root in BI action; in \GA\ 
the coefficient of the $T^2$
term in $k(T)$ was found 
to be $ 3 \ov 32$ in the electric ($q=\bar q=1$) case. 
%This  result suggests that
%the exact expression for this function in the electric case is 
%$k(T) = e^{{1\ov 4} T}$.
}
 Also, if one uses the annulus 
calculations to find the net potential between two such parallel
branes by adding up the potentials between constituents, one finds 
that it vanishes exactly by
the abstruse identity. Thus, the self-dual D3-branes behave very much
like BPS states even though the type 0 theory has no supersymmetry.

The fermionic nature of the strings connecting
an electric and a magnetic D3-brane is crucial for this cancellation.
The normalization of the fermions in the annulus diagram
connecting an electric and a magnetic D3-brane
is the same as between
two self-dual D3-branes of type IIB theory. 
In that case the effective field theory is the
${\cal N}=4$ supersymmetric $U(2)$ broken to
$U(1) \times U(1)$. Each adjoint of $U(2)$ produces fields with charges
$(1, -1)$ and $(-1, 1)$ under the $U(1) \times U(1)$.
Since there are 4 adjoint Weyl fermions in the
${\cal N}=4$ supermultiplet, we find 4 bifundamental fermions
of each type. Generalization to $N>1$ is straightforward: the field theory on
$N$ electric and $N$ magnetic D3-branes of type 0B theory contains
4 Weyl fermions in the
$({\bf N}, \overline {\bf N})$ of $U(N)\times U(N)$ and
4 Weyl fermions in the
$(\overline {\bf N}, {\bf N})$.

The theory with the field content derived above is
quite special. First, it has equal numbers of bosonic and
fermionic degrees of freedom, $16 N^2$, and thus no induced   one-loop
cosmological constant or quartic divergence. One also finds 
that the two coefficients in the (one-loop) gravitational  trace anomaly
are equal, i.e. there is a single central charge as in the $\N=4$
 SYM theory. As we explain below, the theory is actually a
supersymmetry breaking projection of the
$\N=4$  SYM theory.

We assume as usual that
the $U(1)$ factors decouple in the infrared
and consider the $SU(N)\times SU(N)$
theory with the field content derived above.\foot{Requiring the
bosons to belong to the adjoint representation of $SU(N)\times SU(N)$
while keeping fermions in the bifundamental representations 
now implies that the number
of  bosonic and fermionic degrees of freedom match only 
approximately, in the large $N$ limit.}
Based on our conjecture that this field theory is dual to the
$AdS_5\times S^5$ background of type 0B theory, we expect that it is conformal.
Indeed, it is not hard to show that the one-loop beta function
cancels. The  one-loop coefficient $b_1$ in a 
gauge theory with group $G$ coupled
to scalars  transforming in a representation $S$ 
and  (2-component) Weyl fermions transforming in a representation $F$
is
\eqn\twol{
b_1 ={ 11\ov 3}  C_2(G) -   { 1\ov 6 } T_2(S) -
{2\ov 3} T_2 (F)    \ , }
where $T_2$ is 
 the Dynkin index of the  representation, 
$\Tr (T_A T_B) = T_2 \delta_{AB}$. For the adjoint representation of $SU(N)$
$T_2= C_2(G)=N,$
while for the fundamental representation $T_2 = 1/2$.
As far as each $SU(N)$ is concerned, we have 6 adjoint
scalars and $8N$ fundamental Weyl fermions. It follows from
\twol\ that the 1-loop beta function vanishes for each of
the two gauge couplings. In fact, we will set the two
gauge couplings equal.

%%%%%%%%%%%%%%%%%%%%%%%%%%%%%%%%%%%%%%%%%%%%%%%%%
The two-loop  gauge coupling  beta function coefficient 
is \refs{\jones,\MV}
$$
b_2 = {34\ov 3 } [C_2(G)]^2 -  \sum [2C_2(S) + { 1\ov 3}
 C_2(G)] T_2 (S)  $$ 
 \eqn\twl{  -\  2\sum  [C_2(F) + {5\ov 3}  C_2(G)] T_2 (F)
+   Y_4 \ .  }
Here 
 $C_2(R)$ is the eigenvalue of the 
quadratic Casimir operator $\hat C_2 = T_A T_A $  in  
a representation $R$, i.e. 
$ d(R) C_2 (R)=d(G) T_2 (R)$, 
where $d(R)$ and $d(G)$ are  the dimensions  of the 
representation and the group. For the fundamental representation of $SU(N)$\ 
$  C_2(F) = { N^2-1 \ov 2N}$.
\ $Y_4$ is the contribution of the Yukawa couplings \MV, 
\eqn\yuk{
Y_4  = { 1 \ov d(G)} \Tr [ \hat C_2 (F)  Y_I Y^\dagger_I]
\ , }
where $Y_I$ is the matrix determining the   Yukawa 
terms in the Lagrangian, 
$ \Psi \zeta Y_I \Psi \X^I + h.c.$,  
where $\zeta= i\sigma_2$;  $\Psi$ is the full set of 
2-component fermions  and $\X$ is  the  set of scalars. 
We  are assuming  that,  as in the $\N=4$ SYM theory,  the 
Yukawa couplings are equal to the gauge coupling constant. 

The sums in \twl\ are over irreducible representations. 
Since the gauge group 
here is not simple and the  fermions are not singlets in both of the groups
(and thus `intertwine' the renormalizations of the two  gauge couplings),
there is an important subtlety.
In general, for a product group with couplings $g_1$ and $g_2$
one should make make the replacement \MV:\ 
$g^4 C_2(F) T_2(F) \to \sum_{k,l=1,2}\  g^2_k g^2_l 
 C^{(k)}_2 (F) T_2^{(l)} (F)$. 
Since we set $g_1=g_2=g$, the contribution 
of the term  $-2\sum   C_2 (F) T_2(F)$  in \twl\ is thus effectively
doubled, 
i.e. it is $ 2 \times  2 { N^2-1 \ov 2N} \times { 1 \ov 2} \times 8 N$. 
As a result,  
\eqn\fef{
b_2 =  8 - 24 N^2  + Y_4 \ . }
This  may be compared  with the  2-loop 
coefficient in the $\N=4$ SYM theory \refs{\jones,\PW}, 
where the Yukawa contribution is important for the cancellation, 
$b_2 = - 24 N^2  + Y_4 = 0$.  

To compute $Y_4$  in our case 
 let us specify the structure of the interaction terms
in the action 
more explicitly.  We shall assume that, like the $\N=4$ SYM theory,  
our theory has a global $SU(4)$ symmetry. 
The fermions $\Psi=(\psi, \psi')$ in the representations 
$(\overline {\bf N}, {\bf N})$   and 
$( {\bf N}, \overline {\bf N})$ each transform as the $\bf 4$ of $SU(4)$
(and  their hermitian conjugates as the $\bar {\bf 4}$), 
while the adjoint scalars $\X=(X,X')$  each transform as
the $\bf 6$ of $SU(4)$.
Then the couplings of the fermions to
the $SU(N) \times SU(N)$ gauge field $(A,A')$ are, symbolically, 
$A (\psi^\dagger  \psi + \psi'  \psi'^\dagger)$
and $A'(\psi'^\dagger  \psi' + \psi \psi^\dagger)$, while
the Yukawa couplings are  
$X_{ab} \psi^a  \zeta  \psi'^b + h.c.$ and 
$X'_{ab} \psi^a\zeta  \psi'^b + h.c.$,  where $a,b=1,2,3,4$.
%As a result, the $SU(4)$ part of the Yukawa matrix $Y$ is the same 
%as in the $\N=4$ SYM, $\d^a_c \d^b_d - \d^a_d \d^b_c$, 
%and the trace of its square is  again $24$. 
%Finally, $Y_4 = 24 N^2$ and thus 
%$b_2 =  8 - 24 N^2  + 24 N^2$, i.e. 
%our theory is 2-loop finite in the large $N$ limit.
The value of the Yukawa contribution  $Y_4$ then turns out to be\foot{The
$SU(4)$ part of the Yukawa matrix $Y$ is the same as in the $\N=4$ SYM,
$\d^a_c \d^b_d - \d^a_d \d^b_c$,
and the trace of its square is  again $24$.
Note  that all the Yukawa couplings (involving both sets of the 
scalar fields) contribute to the renormalization of each of the gauge 
coupling. Another factor of 2 is related to the `off-diagonal' structure
of the Yukawa matrix.}  
\eqn\yyy{
Y_4 =  24 \times 2 \times 2 N \times{d(F)\over d(G)} [C_2(F)]^2 =
96  C_2(F) T_2 (F) =  24
(N^2-1)
\ .
}
As a result, the leading $O(N^2)$ term in \fef\ 
cancels out and we have $b_2=-16$.\foot{This is similar to what happens
in the `orbifold' field theories of \refs{\KS,\LNV,\ber}
as discussed explicitly  at the 2-loop level in \FR.}
Thus, the theory is 2-loop finite in the large $N$ limit.

%%%%%%%%%%%%%%%%%%%%%%%%%%%%%%%%%%%%%%%%%%%%%%%%%%%%%%%%
In fact, we believe that the non-supersymmetric
$U(N)\times U(N)$ theory we are considering can obtained
from the ${\cal N}=4$ supersymmetric  $U(2N)$
SYM theory  by twisting it with $(-1)^F$ in a very similar way to
how the orbifold $Z_2$ theory is obtained \dm. The
arguments of \ber\ may then explain why this theory is conformal
in the planar limit.

In the orbifold case of \dm\ the $Z_2$ 
reverses the sign of four of the scalar fields $X$
and also of their fermionic partners. If we think of the 
${\cal N}=4$ multiplet
as an $\N=2$  vector multiplet plus a hypermultiplet, 
then the $Z_2$ reverses the sign of the hypermultiplet fields.
Now, break up the $2N \times 2N$ matrices into four $N\times N$ blocks.
On the diagonal blocks we keep the fields invariant under the $Z_2$,
i.e. the vector multiplets. On the off-diagonal blocks we keep the
non-invariant fields, i.e. the hypermultiplets.
This produces the $\N=2$ supersymmetric $U(N)\times U(N)$ theory
coupled to two bifundamental hypermultiplets \dm.
In effect, we are keeping the fields invariant under the change of sign
of the hypermultiplet accompanied by the $U(2N)$
gauge transformation $\II$ which is the conjugation by
$\pmatrix{ I&  0\cr  0 & -I\cr}$, where 
$I$ is the $N \times N$ identity matrix.

In our case we take the $Z_2$ action to be $\II\cdot (-1)^F$, 
i.e. the $U(2N)$ gauge transformation is accompanied by 
the change of sign of all the fermionic fields.
Now on the diagonal blocks we keep the invariant fields: the
gauge field and the 6 scalars. On the off-diagonal blocks we keep
the 4 Weyl fermions. This gives the $U(N)\times U(N)$
theory with 6 scalar
adjoints of each $U(N)$, and 8 bifundamental Weyl fermions.
The meaning of the $\II$ transformation $\pmatrix{ A&  B\cr  C &  D\cr}
\to  \pmatrix{ A&  -B\cr  -C &  D\cr} $ 
is that it projects out the states corresponding to 
open strings connecting $N$ electric and $N$ magnetic 
branes; combining it with $(-1)^F$ then allows back the fermionic
connecting strings.

%%%%%%%%%%%%%%%%%%%%%%%%%%%%%%%%%%%
\newsec{Discussion}
%%%%%%%%%%%%%%%%%%%%%%%%%%%%%

One fascinating application of the AdS/CFT correspondence is
towards the classification of 4-dimensional conformal gauge theories. 
For example, interesting large $N$
CFT's with less than maximal supersymmetry
are found in the type IIB setting when the $S^5$ is replaced by
orbifolds of $S^5$ \refs{\KS,\LNV} or by other Einstein manifolds \KW.
Certain non-supersymmetric large $N$ CFT's whose field theoretic
formulation is less transparent were studied in \refs{\Gir}.
In this paper we propose another route towards constructing
{\it non}-supersymmetric large $N$ CFT's: by looking for dual backgrounds
of the type 0B theory. The simplest, and most symmetric, background
is $AdS_5\times S^5$.  Here we have found  the dual gauge theory
which implements the corresponding $SO(2,4)\times SO(6)$ symmetry.
Just like most non-supersymmetric orbifold CFT's
\refs{\KS,\LNV,\FR}, this theory
appears to be conformal only in the large $N$
`t Hooft limit. If  $N$ is kept finite then the 2-loop beta function does
not cancel exactly. 

We have shown that our CFT may be obtained by a
$\II \cdot (-1)^F$ projection of the $\N=4$ SYM theory.
The $(-1)^F$ projection  is related to the general fact that the type 0 theory
is a $(-1)^F$ orbifold ($F$ is the full space-time fermion number)
 of the type II theory \refs{\DH,\AW,\DS,\DB}.
 Indeed, the space-time fermions, which are the non-invariant states, are
projected out. 
Thus, it is not too  surprising that the self-dual
3-brane of type 0B theory is related by a similar projection
to the self-dual 3-brane of type IIB theory.
As we have seen, the full $Z_2$ action on the stack of
D3-branes  includes also the gauge transformation $\II$.

We have argued  that the $AdS_5 \times S^5$
background, which exists only for a non-vanishing R-R flux, is stable
when its radius is of order the string scale or smaller.
This suggests that the type 0 theory  is  not just  a  formal   
twisting of  type II theory but, in fact, fits naturally into the
AdS/CFT duality. 
%natural element of  a bigger  picture.
Indeed, there is a general relation between type II and
type 0 theories which may have new
implications in the presence of a (large) number of D-branes or R-R flux.
%than in pure flat space  vacuum case previously discussed.
As was  pointed out  in \refs{\DS,\DB},
compactifying type II theory on a  9-circle  and
 applying  the $(-1)^F$ projection 
in combination with $(-1)^{ 2\pi i R_9 P_9}$
 ($P_9$ is momentum operator) 
gives a theory which interpolates between the type IIB 
theory at $R_9 \to \infty $ and the type 0B theory at $R_9 \to  0$.
%since the twisting leads to the appearance
%of tachyon at certain radius.
Equivalently, considering type II theory  at finite temperature, i.e. 
compactifying the  euclidean time direction with antiperiodic 
boundary conditions for space-time fermions, 
one concludes \AW\ that the type 0 theory is the high temperature
limit of the type II theory. 
Of course, in flat space one expects a Hagedorn phase 
transition to take place at the temperature of order the string
scale, so that the high-temperature  limit is  formal.
It may be interesting, however, to reexamine the issue of the phase
transition for backgrounds with R-R flux. 
% identified
%with the formal  large or infinite temperature limit of type II theory.
%Increasing the temperature one finds that 
%the interpolating theory develops a tachyon  at the radius of
% $X^0$  smaller than corresponding to the Hagedorn temperature.
%As  suggested   in \AW, one of course expects 
%that at the Hagedorn temperature  corresponds to  a phase transition, 
%i.e. one  does not expect to   get a `physical' tachyon state.
%It might be of interest 
%to reconsider the  discussion of the phase transition in \AW\
%(in particular, the winding-mode tachyon  -- dilaton 
%effective action) 
%which was in flat space 
%in the presence of  a large  background R-R flux.

The relation between the type 0 theory and the thermal type II
theory suggests a connection between the type 0 scenario
for constructing non-supersymmetric gauge theory and the temperature
induced supersymmetry breaking proposed by Witten \newWit.
However, in \newWit\ the temperature was taken to be much lower
than the string scale while in the type 0 setting it is much higher.
Nevertheless, there are some similarities between the two approaches
which may be pursued further.

\bigskip

{\it Note added:}
\medskip

Using the general expressions \refs{\OO,\JO}\ for 1-loop renormalization 
of the scalar potential and the Yukawa    matrix 
(in the form given in  eq. (3.45) in \JO) 
we have  checked that  the quartic scalar coupling and the Yukawa 
coupling of the non-supersymmetric 
$SU(N) \times SU(N)$   gauge theory
described above 
 are not renormalized  at  one loop
to the leading order in large $N$.\foot{We are grateful to E. Witten 
for raising the question about the
scalar and Yukawa coupling renormalizations in this theory.} 
In contrast to the 1-loop  gauge coupling beta-function (but similar to the  
2-loop one),  
  the  1-loop  scalar and Yukawa beta-functions  contain  non-vanishing
subleading terms. 
We expect  a similar pattern of 1-loop renormalizations in 
other large $N$ CFT's \refs{\KS,\LNV} 
 obtained by `orbifolding' the $\N=4$ SYM theory.

As in the $\N=4$ SYM theory (but in contrast to  the `electric' 
theory \KT\ which has the same bosonic 
spectrum but no fermions) 
here the fermionic contribution  cancels 
not only the $\tr([X,X]^2)$ 
term  but also the (absent in the classical action)
term $\tr (X^2 X^2)$ 
in the divergent part of the 1-loop effective action.
However, in contrast to the $\N=4$ SYM  here 
these  cancellations happen 
only to leading order in the large $N$ expansion.\foot{The reasons for 
the large $N$  cancellation are the same as in the 1-loop gauge 
coupling beta-function:  (i) the traces in 
the adjoint  and fundamental representations $\Tr$ and $\tr$ 
 are related,  for large $N$, simply 
by the factor of $2N$, and (ii)  while in the $\N=4$ SYM 
one has 4  adjoint fermions,  here we have $4\times 2N$ types  of  fermions 
 in the  fundamental  representation of $SU(N)$.
Since in general for a matrix $X$ from the $SU(N)$ algebra   \ 
$\Tr X^2 = 2 N \tr X^2$ but 
$\Tr X^4 = 2N \tr X^4 + 6 \tr X^2 \tr X^2$,
%there is still a  non-vanishing 
%subleading (`non-planar') contribution to the 
%renormalization of the 
the quantum effects induce certain trace-squared terms 
in the scalar potential, but the renormalization of single-trace
terms cancels in the large $N$ limit.
%Thus at finite $N$ one needs to add additional  counterterms 
%to the classical action.
}
A similar conclusion is reached for the Yukawa coupling renormalization.

The reason for the  breaking  of conformal  invariance
at subleading orders in $1/N$  can be understood 
from the  string-theory point of view.\foot{We are grateful to 
E. Silverstein for pointing this out to us.}  
Type 0 string theory has a non-vanishing 
partition function on the torus \DH, implying the presence 
of the  string-theory 1-loop cosmological constant
in the effective action. The tree-level 
Einstein frame action $S_0$ \fii\  should be  supplemented by
the 1-loop one\ 
$S_1 = c_1 \int d^{10} x  \sqrt { -g}  \  e^{{5\ov 2} \Phi} \ w(T)$, 
where $w(T) = 1 + a_1 T + ...$ is a function of the tachyon field.
In contrast to what happens in the type IIB theory,  where $c_1=0$
because of the underlying supersymmetry,  
here the  field equations  that follow
from $S=S_0 +S_1$   no longer have $\P=\const,\ T=0$ 
and thus $AdS_5 \times S^5$ is no longer an exact solution.
Thus, as in the electric theory case \KTnew, 
one should  find  a non-trivial RG flow in the 
corresponding world-volume theory. 
In the present self-dual case the flow is thus  suppressed by $1/N^2$. 

\bigskip

{\it Note added in proof:}
\medskip
After this paper was completed,  Nekrasov and
Shatashvili pointed out  \NS\  that the $Z_2$ operation
$(-1)^F$ which changes the sign of all fermion fields is simply the
$-1$ from the center of the $SU(4)$ R-symmetry group. 
Thus our theory may be viewed as
another orbifold of the ${\cal N}=4$ SYM theory
by a discrete subgroup of the R-symmetry. In view of the results in
\ber, this strengthens the argument that this theory is exactly 
conformal in the planar limit.
A special feature of this particular orbifold field theory is that
it cannot be realized on D3-branes of type IIB theory placed at an
orbifold singularity of $R^6/\Gamma$.

%%%%%%%%%%%%%%%%%%%%%%%%%%%%%%%%%%%%%%%%%%%%%%%%%
\newsec{Acknowledgements}
%%%%%%%%%%%%%%%%%%%%%%%%%%%%%%%
We are grateful to  E. Kiritsis, I. Kogan, H. Liu,
V. Periwal, E. Silverstein and E. Witten for useful discussions
and comments.
The work  of I.R.K. was supported in part by the NSF
grant PHY-9802484 and
by the James S. McDonnell
Foundation Grant No. 91-48.
The  work  of A.A.T. was supported in part
by PPARC, the European
Commission TMR programme grant ERBFMRX-CT96-0045
and  the INTAS grant No.96-538.
%%%%%%%%%%%%%%%%%%%%%%%%%%%%%%%%%%%%%%%%%%%%%%%%%%%%%%%%
\vfill\eject
\listrefs
\end